\documentclass[journal]{IEEEtran}
\IEEEoverridecommandlockouts
\usepackage{cite}
\usepackage{amsmath,amssymb,amsfonts}
\usepackage{algorithmic}
\usepackage{graphicx}
\usepackage{textcomp}
\usepackage{xcolor}
\usepackage{amsfonts}
\usepackage{bm}
\usepackage{booktabs}
\usepackage{algorithm}
\usepackage{multirow,bm,bbm,array,setspace}
\usepackage{textcomp}
\usepackage{psfrag}
\usepackage{pstricks,enumerate}
\usepackage{bbm,subfigure}
\usepackage{dblfloatfix}

\makeatletter
\newcommand{\distas}[1]{\mathbin{\overset{#1}{\kern\z@\sim}}}%
\newsavebox{\mybox}\newsavebox{\mysim}
\newcommand{\distras}[1]{%
  \savebox{\mybox}{\hbox{\kern1pt$\scriptstyle#1$\kern1pt}}%
  \savebox{\mysim}{\hbox{$\sim$}}%
  \mathbin{\overset{#1}{\kern\z@\resizebox{\wd\mybox}{\ht\mysim}{$\sim$}}}%
}
\makeatother
\ifCLASSINFOpdf
\else
\fi
\newcommand{\bH}{\bm{H}}

\newcommand{\bI}{\bm{I}}

\newcommand{\bJ}{\bm{J}}

\newcommand{\bV}{\bm V}

\newcommand{\bY}{\bm{Y}}

\newcommand{\bQ}{\bm{Q}}

\newcommand{\bA}{\bm{A}}

\newcommand{\bG}{\bm{G}}

\newcommand{\bF}{\bm{F}}

\newcommand{\bD}{\bm{D}}
\newcommand{\bGa}{\bm{\Gamma}}
\newcommand{\hh}{\mathrm{H}}
\newcommand{\tr}{\mathrm{tr}}

\newtheorem{proposition}{Proposition}
\newtheorem{lemma}{Lemma}
\newtheorem{theorem}{Theorem}

\newtheorem{remark}{Remark}

\begin{document}

\title{Finite Horizon Optimization for Large-Scale MIMO}
\author{
\IEEEauthorblockN{
    Yi Feng, \IEEEmembership{Graduate Student Member,~IEEE} and Kaiming Shen, \IEEEmembership{Senior Member,~IEEE}
}
\thanks{
Manuscript submitted on \today.

The authors are with School of Science and Engineering, The Chinese University of Hong Kong, Shenzhen, China (e-mails: yifeng1@link.cuhk.edu.cn; shenkaiming@cuhk.edu.cn).}
} % <-this % stops a space

\maketitle

\begin{abstract}
Large-scale multiple-input multiple-output (MIMO) is an emerging wireless technology that deploys thousands of transmit antennas at the base-station to boost spectral efficiency. The classic weighted minimum mean-square-error (WMMSE) algorithm for beamforming is no suited for the large-scale MIMO because each iteration of the algorithm then requires inverting a matrix whose size equals the number of transmit antennas. While the existing methods such as the reduced WMMSE algorithm seek to decrease the size of matrix to invert, this work proposes to eliminate this large matrix inversion completely by applying gradient descent method in conjunction with fractional programming. Furthermore, we optimize the step sizes for gradient descent from a finite horizon optimization perspective, aiming to maximize the performance after a limited number of iterations of gradient descent. Simulations show that the proposed algorithm is much more efficient than the WMMSE algorithm in optimizing the large-scale MIMO precoders.
\end{abstract}


\section{Introduction}
\begin{figure}[t]
    \centering
    \includegraphics[width=1\linewidth]{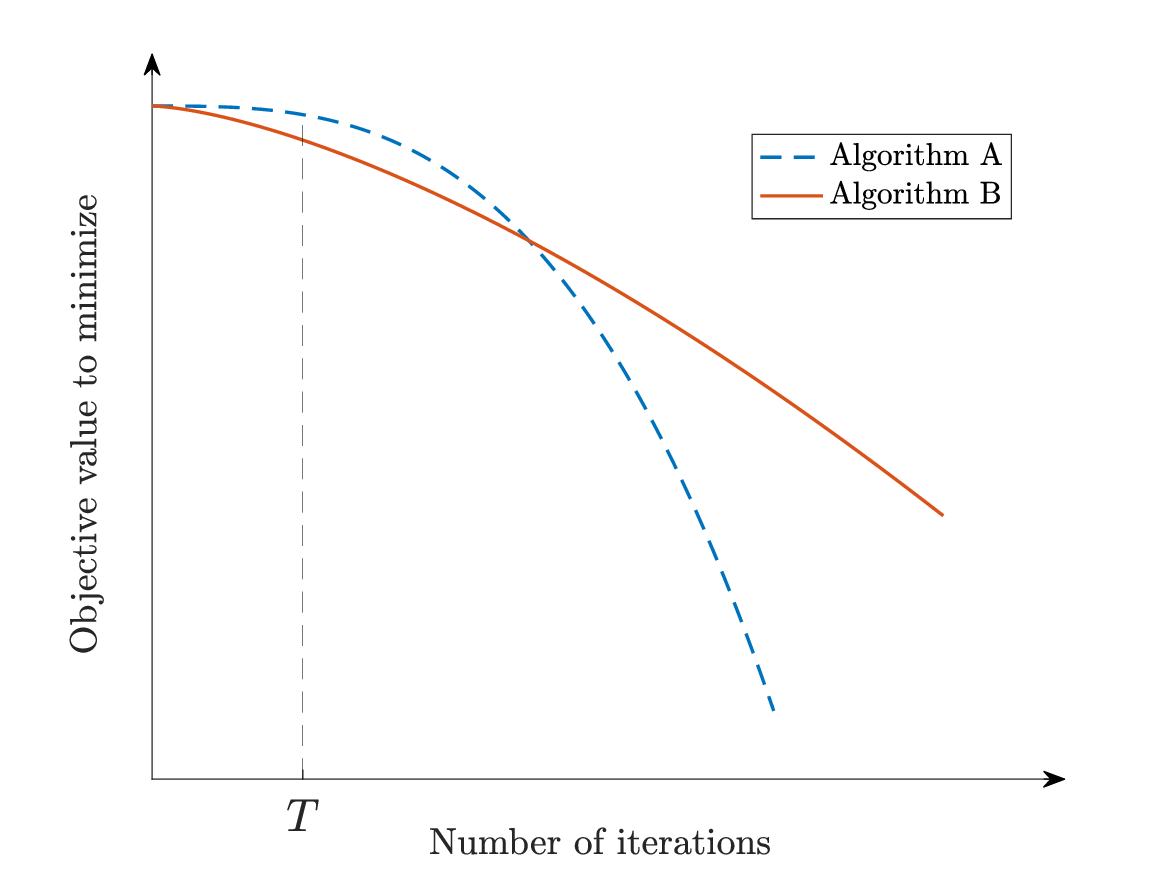}
    \caption{According to the conventional optimization theory, algorithm A
    is better because its asymptotic convergence rate is higher when the number of iterations is sufficiently large. However, in the context of finite horizon optimization, algorithm B can outperform algorithm A when the attention is restricted to a finite range up to $T$ iterations.}
    \label{figdraw}
\end{figure}

\IEEEPARstart{L}{arge-scale} multiple-input multiple-output (MIMO) has imposed a huge challenge to the wireless network optimization because the corresponding beamforming problem is not only nonconvex but also involves high-dimensional variables \cite{liu2024survey}. Unfortunately, the well-known weighted minimum mean-square-error (WMMSE) algorithm \cite{christensen2008weighted,shi2011iteratively} is not suited for the large-scale MIMO case, because it entails inverting a large matrix whose size equals the number of transmit antennas for each iteration. To remedy the situation, this work examines the large-scale MIMO beamforming problem from a novel viewpoint of \emph{finite horizon optimization}: how to maximize the performance given a limited number of iterations as illustrated in Fig.~\ref{figdraw}.

%The massive multiple-input multiple-output (MIMO) network is a promising technology that can provide high throughput, reliability, and energy efficiency. A crucial problem in this field is designing transmit beamformers to maximize the system weighted sum-rate (WSR) subject to power budget. Unfortunately, the general WSRMax problem is not yet amendable to a convex formulation. In fact, it is NP-hard \cite{luo2008dynamic}. This paper considers solving WSRMax problem by using Finite Horizon optimization, which focuses on optimizing the algorithm under a fixed iteration budget $T$. 

For the beamforming problem, many early works \cite{yu2007transmitter,ng2008distributed,rashid1998transmit,zakhour2013min,dahrouj2010coordinated} pursue the minimum total power of precoders under the given signal-to-interference-plus-noise ratio (SINR) constraints. Despite the nonconvexity, this min-power beamforming problem can be solved globally in an efficient way based on the uplink-downlink duality \cite{yu2007transmitter}. Another beamforming problem formulation in the literature \cite{chang2008approximation,zhang2013semidefinite,ashraphijuo2017multicast} is to maximize the minimum SINR across receivers under the given power constraint; the resulting max-min problem is amenable to the semidefinite programming (SDP) approach. In particular, the max-min SINR problem can be approximately solved to within a constant factor for the single-cell network \cite{chang2008approximation}. The modern works in the area mostly study the weighted sum rates (WSR) maximization for the beamforming problem. Since the WSR problem is NP-hard \cite{luo2008dynamic}, the previous works that aim at global optimum must resort to those exponential-time optimization methods, e.g., the exhaustive search \cite{cendrillon2006optimal}, the branch-and-bound algorithm \cite{cumanan2010joint}, and the polyblock optimization \cite{qian2009mapel}. Some other existing works have suggested various approximations of the WSR problem to make the solution practical, e.g., the signomial programming approximation \cite{boyd2007tutorial}, the geometric programming approximation  \cite{chiang2005geometric}, the successive convex approximation \cite{kim2011optimal}, and the minorization-maximization (MM) approximation \cite{gong2020majorization}. In particular, as a classic and simple approximation, the zero-forcing (ZF) beamforming method \cite{goldsmith2005wireless} just ignores the background noise and then nullifies the interfering signals. The authors of \cite{sun2010eigen} improve the ZF beamforming method to reduce the approximation error, by dropping a set of weakest eigen-channels. Moreover, it is worth mentioning the information theoretic approach to the beamforming problem based on interference alignment \cite{cadambe2008interference,maddah2008communication,razaviyayn2011degrees,sridharan2016role}, wherein the data rate in the WSR problem is approximated as  the degrees-of-freedom (DoF).

The WMMSE algorithm \cite{christensen2008weighted,shi2011iteratively}, first proposed about 20 years ago, still constitutes one of the most popular beamforming methods nowadays. It derives from the following insight: the data rate maximization is equivalent to the received signal mean-square-error (MSE) minimization. Furthermore, \cite{shen2018fractional1} shows that the WMMSE algorithm is intimately related to fractional programming (FP). From an FP perspective, the WMMSE algorithm can be interpreted as a two-stage method: first, it moves the SINR terms to the outside of $\log(1+\text{SINR})$ to obtain a sum-of-ratios problem; second, it addresses the sum-of-ratios problem by the quadratic transform (QT) \cite{shen2018fractional1}. We remark that, at the second stage, the QT can be performed in various ways, one particular way leading to the WMMSE algorithm. Thus, the WMMSE algorithm turns out to be a special case of FP, and it may not be the best case of FP for beamforming as shown in \cite{shen2018fractional2,shen2019optimization}.

The WMMSE algorithm \cite{christensen2008weighted,shi2011iteratively} is based on an iterative paradigm. As a desirable feature, WMMSE provides closed-form solution for each iteration. Note that such closed-form solution can be interpreted as the projection onto the surface of an $M$-dimensional ellipsoid \cite{shen2024accelerating}, where $M$ is the number of transmit antennas, so it requires inverting an $M\times M$ matrix. This matrix inverse operation becomes costly when $M$ is large, so the WMMSE algorithm is not suited for the large-scale MIMO beamforming (e.g., with thousands of antennas deployed \cite{de2020non,de2021quasi}).
To address this issue, a recent work \cite{zhao2023rethinking} proposes the so-called \emph{reduced WMMSE} algorithm that maps the original beamforming variable from the $M\times M$ space into the $NK\times NK $ space, where $N$ is the number of receive antennas at each user terminal and $K$ is the number of user terminals per cell. As a consequence, the new algorithm requires inverting an $NK\times NK$ matrix instead for each iteration. Yet there are two subtle issues with the reduced WMMSE algorithm \cite{zhao2023rethinking}. First, it is possible that $NK>M$ in practice, and then the $NK\times NK$ matrix inversion can be even more complex than the $M\times M$ one. Second, it can be difficult to recover the original solution for the multi-cell network, because the beamforming variable mapping is shown to be invertible only for the single-cell case. 

Alternatively, a more recent work \cite{zhang2023enhancing} proposes to eliminate the matrix inversion by the \emph{nonhomogeneous quadratic transform (NQT)}---an improved version of the QT method for FP \cite{shen2018fractional1}. However, the NQT would slow down the convergence of the iterative optimization. As shown in \cite{shen2024accelerating}, both NQT and WMMSE can be interpreted as constructing a concave surrogate function for the original objective function of the beamforming problem, but the surrogate function constructed by WMMSE is tighter, so the WMMSE algorithm \cite{christensen2008weighted,shi2011iteratively} converges faster than the NQT algorithm \cite{zhang2023enhancing}. To be more specific, letting $f^{t}$ be the value of the WSR objective after $t$ iterations, and $f^\star$ the value of convergence, \cite{shen2024accelerating} shows that WMMSE achieves
\begin{equation}
    |f^{t}-f^\star|= \frac{\alpha}{t}\quad\text{as}\quad t\rightarrow\infty,
\end{equation}
while the QT method achieves
\begin{equation}
    |f^{t}-f^\star|= \frac{\beta}{t}\quad\text{as}\quad t\rightarrow\infty,
\end{equation}
where $\beta>\alpha>0$. Furthermore, \cite{shen2024accelerating} shows that the NQT method can be interpreted as a gradient projection, so Nesterov's extrapolation scheme \cite{nesterov1983method} can be incorporated into NQT to accelerate its convergence. Eventually, \cite{shen2024accelerating} shows that the extrapolated NQT leads to the following convergence rate:
\begin{equation}
\label{asymptotic rate 2}
    |f^{t}-f^\star|= \frac{\gamma}{t^2}\quad\text{as}\quad t\rightarrow\infty,
\end{equation}
where $\gamma>0$. Thus, the extrapolated NQT has much faster convergence than WMMSE and NQT.

Nevertheless, the above convergence rate analyses from \cite{shen2024accelerating} only reflect the asymptotic speed as $t\rightarrow\infty$. But for a real-world network it is the finite or even small $t$ that truly matters. Toward this end, this work considers the \emph{finite horizon optimization} of the large-scale MIMO precoders. Specifically, given a finite positive integer $T$, we aim to maximize the performance after $T$ iterations, i.e.,
\begin{equation}
    \mathrm{maximize}\quad f^T\quad\text{for some given}\quad T<\infty.
\end{equation}
The finite horizon optimization is a newly emerging notion initiated in the machine learning field \cite{zhang2024finite} to facilitate the finite-depth neural network training \cite{gregor2010learning,sun2016deep}. To the best of our knowledge, the finite horizon optimization has not yet been explored in the area of wireless communications and signal processing. We wish to clarify the distinction between the finite horizon optimization and the optimization with finite convergence (e.g., the conjugate gradient descent \cite{shanno1978convergence}). 
The finite horizon optimization does not guarantee convergence to a global optimum or stationary point after $T$ iterations. Rather, it aims at 
what one can expect at best after $T$ iterations. The main contributions of this paper are summarized below:
\begin{itemize}
\item We recast the constrained log-det beamforming problem into an unconstrained quadratic program by means of the matrix FP, extending the analogous result by Zhao, Lu, Shi, and Luo \cite{zhao2023rethinking}.
\item We propose solving the unconstrained quadratic program by the finite-horizon gradient descent, and thereby avoid inverting a large matrix whose size equals the number of transmit antennas of the large-scale MIMO.
\item We further optimize the step sizes of the finite-horizon gradient descent by the Chebyshev polynomial theory. To the best of our knowledge, this is the first time that the finite horizon optimization is applied to the communication system design.
\end{itemize}

\emph{Notation:} For a matrix $\bA$, $\bA^\hh$ denotes its conjugate transpose, $\|\bA\|_2$ denotes its $\ell_2$ norm that equals the largest singular value of $\bA$, $\|\bA\|_F$ denotes its Frobenius norm. For a square matrix $\bA$, $\tr(\bA)$ denotes its trace, and $|\bA|$ denotes its determinate. For a nonsingular matrix $\bA$, $\bA^{-1}$ denotes its inverse. Moreover, $\bI$ denotes the identity matrix, $\mathbb C^{d}$ the set of $d \times1$ vectors, $\mathbb C^{d\times m}$ the set of $d\times m$ matrices, and $\mathbb S_{+}$ the set of positive semidefinite matrices. For a complex number $a\in\mathbb C$, $\Re\{a\}$ is its real part, $|a|$ is its absolute value. For a differentiable function $g(x)$, $\nabla g(x)$ denotes its gradient. Let $\mathbb P_T=\{c_Tx^T+c_{T-1}x^{T-1}+\ldots+c_1x+c_0:c_T\ne0\}$ be the set of degree-$T$ polynomials given a positive integer $T$.

\section{System Model}

Consider a single-cell downlink large-scale MIMO network; the multi-cell case is discussed later in Section \uppercase\expandafter{\romannumeral5}. Assume that the base-station (BS) has $M$ transmit antennas and each user terminal has $N$ antennas; it shall be understood that $M\gg N$ in the large-scale MIMO system. The BS aims to send independent messages to a total of $K$ user terminals in the cell simultaneously via spatial multiplexing. Assume that at most $d\le N$ data streams can be transmitted to the same user terminal. We use $k,j=1,\ldots,K$ to index the users. Denote by $\bH_{k}\in\mathbb C^{N\times M}$ the downlink channel from BS to the $k$th user, $\bV_{k}\in\mathbb C^{M \times d}$ the transmit precoder intended for the $k$th user, and $\sigma^2$ the background noise power.
Thus, an achievable rate $R_k$ of user terminal $k$ can be computed as
\begin{equation}
\label{R_k}
    R_k = \log\left|\bI+\bV_{k}^\hh\bH_{k}^\hh\bF_{k}^{-1}
\bH_{k}\bV_{k}\right|,
\end{equation}
where
\begin{equation}
    \label{F}
   \bF_{k} = \sigma^2\bI+\sum_{j\ne k}\bH_{k}\bV_{j}\bV_{j}^\hh\bH_{k}^\hh.
\end{equation}
Each user terminal $k$ is assigned a positive weight $\omega_{k}>0$ in accordance with its priority. We seek the optimal precoding variable $\bV=\{\bV_{k}\}$ to maximize the WSR:
\begin{subequations}
\label{prob:MIMO}
\begin{align}
\underset{\bV}{\text{maximize}} &\quad \sum^K_{k=1}\omega_{k}R_k
\label{prob:MIMO:obj}\\
  \text{subject to} & \quad \sum^K_{k=1}\|\bV_{k}\|^2_F\leq P,
  \label{prob:MIMO:constraint}
\end{align}
\end{subequations}
where $P$ is the power constraint. We remark that the number of transmit antennas $M$ in \eqref{prob:MIMO} is much larger than the conventional MIMO case, e.g., it is of the order of thousands.

\section{Conversion to Quadratic Program}
\label{sec:QP}

It is difficult to tackle the problem \eqref{prob:MIMO} directly because of its nonconvexity. This section shows that the problem can be rewritten as an unconstrained quadratic program.
For the conventional MIMO network, the new problem is favorable in that each $\bV_k$ can then be solved in closed form. Nevertheless, for the large-scale MIMO case with $M$ being large, the closed-form solution of $\bV_k$ entails an $M\times M$ matrix inversion which is computationally formidable in practice. A finite horizon optimization treatment of this new quadratic program is discussed in Section \uppercase\expandafter{\romannumeral4}.

First of all, it is easy to notice that the power constraint \eqref{prob:MIMO:constraint} must be tight at any \emph{nontrivial}\footnote{As defined in \cite{zhao2023rethinking}, the solution $\bV$ is said to be nontrivial if $\bH_k\bV_k\ne0$ for at least one $k=1,2,\ldots,K$.} stationary-point solution, for otherwise we can further increase the value of $\sum^K_{k=1}\omega_kR_k$ by scaling up all the $\bV_{k}$'s simultaneously. Since the global optimum must be nontrivial, problem \eqref{prob:MIMO} is equivalent to
\begin{subequations}
\label{prob:MIMO:tight}
\begin{align}
\underset{\bV}{\text{maximize}} &\quad \sum^K_{k=1}\omega_{k}R_k
\label{prob:MIMO:obj:tight}\\
  \text{subject to} & \quad \sum^K_{k=1}\|\bV_{k}\|^2_F= P,
  \label{prob:MIMO:constraint:tight}
\end{align}
\end{subequations}
in the sense that the two problems have the same global optimum (and also the same set of nontrivial stationary points). Furthermore, problem \eqref{prob:MIMO:tight} can be immediately rewritten as
\begin{subequations}
\label{prob:MIMO:new}
\begin{align}
\underset{\bV}{\text{maximize}} &\quad \sum^K_{k=1}\omega_{k}\log\left|\bI+\bV_{k}^\hh\bH_{k}^\hh\widetilde\bF_{k}^{-1}
\bH_{k}\bV_{k}\right|
\label{prob:MIMO:obj:newsingle}\\
  \text{subject to} & \quad \sum^K_{k=1}\|\bV_{k}\|^2_F= P,
  \label{prob:MIMO:constraint:new}
\end{align}
\end{subequations}
with $\bF_{k}$ in \eqref{F} replaced by
\begin{equation}
    \label{Fq}
   \widetilde\bF_{k} = \Bigg(\frac{\sigma^2}{P}\sum^K_{k=1}\|\bV_{k}\|^2_F\Bigg)\bI+\sum_{j\ne k}\bH_{k}\bV_{j}\bV_{j}^\hh\bH_{k}^\hh.
\end{equation}
We now remove the power equality constraint \eqref{prob:MIMO:constraint:new} and obtain an unconstrained reformulation of problem \eqref{prob:MIMO:new} as
\begin{align}
\underset{\bV}{\text{maximize}} &\quad \sum^K_{k=1}\omega_{k}\log\left|\bI+\bV_{k}^\hh\bH_{k}^\hh\widetilde\bF_{k}^{-1}
\bH_{k}\bV_{k}\right|.
\label{prob:MIMO:unconstrained}
\end{align}
A crucial trait of the above new problem is that its objective value does not alter when all the $\bV_k$ are multiplied by the same nonzero factor, as stated in the following lemma.
\begin{lemma}
\label{lemma:orthogonal}
Denote by $g(\bV)$ the objective function in \eqref{prob:MIMO:unconstrained}. For any nonzero $\bV$, we have $g(\bV)=g(c\bV)$ given any nonzero factor $c$, and thus
\begin{equation}
    \nabla g(\bV) \perp \bV
\end{equation}
for any $\bV$.
\end{lemma}

In light of the above lemma, we can show that\footnote{A weak version of Theorem \ref{theorem:unconstrained} has been shown in \cite{zhao2023rethinking} assuming that $\bV$ is nontrivial. The proof technique in \cite{zhao2023rethinking} is different and is more computationally intensive: it explicitly compares the KKT conditions of \eqref{prob:MIMO} and \eqref{prob:MIMO:unconstrained}.}:
\begin{theorem}
\label{theorem:unconstrained}
    For any $\bV$ with $\sum^K_{k=1}\|\bV_k\|_F^2>0$, it is a stationary point of problem \eqref{prob:MIMO:unconstrained} if and only if $\rho\bV$ is a stationary point of problem \eqref{prob:MIMO}, where 
    $\rho=(P/\sum^K_{k=1}\|\bV_k\|_F^2)^{1/2}$.
\end{theorem}
\begin{IEEEproof}
Denote by $f(\bV)$ the objective function in \eqref{prob:MIMO}. If $\bV$ is a trivial solution so that $\bH_k\bV_k=\bm0$ for every $k=1,2,\ldots,K$, then we have $\nabla f(\bV)=\nabla g(\bV)=\bm 0$, so the theorem holds in this case. The rest of the proof assumes nontrivial $\bV$.

We start with the necessity. If $\bV$ is a stationary point of \eqref{prob:MIMO:unconstrained}, then $\rho\bV$ is a stationary point of \eqref{prob:MIMO:unconstrained} as well because $\nabla g(\rho\bV)=\nabla g(\bV)=\bm0$ according to Lemma \ref{lemma:orthogonal}. Clearly, $\rho\bV$ is feasible for the power constraint and continues to be a stationary point in \eqref{prob:MIMO:new}. Since \eqref{prob:MIMO:new} is equivalent to \eqref{prob:MIMO} in the nontrivial regime, we have $\rho\bV$ be a stationary point of \eqref{prob:MIMO}.

We then show the sufficiency. If $\rho\bV$ is a stationary point of \eqref{prob:MIMO}, then it is also a stationary point of \eqref{prob:MIMO:new} since the two problems are equivalent in the nontrivial regime. Thus, $\rho\bV$ must satisfy the KKT condition in \eqref{prob:MIMO:new}, i.e., $\nabla g(\rho\bV)+2\mu^\star(\rho\bV)=\bm 0$, where $\mu^\star\in\mathbb R$ is the optimal Lagrange multiplier for the constraint $\sum^K_{k=1}\|\bV_{k}\|^2_F= P$. We rewrite this KKT condition as $\nabla g(\bV')=-2\mu^\star\bV'$ where $\bV'=\rho\bV$. Combining the above result with Lemma \ref{lemma:orthogonal}, i.e., $\nabla g(\bV')\perp \bV'$, we obtain $\mu^\star=0$ and $\nabla g(\bV')=\bm 0$. Again, by Lemma \ref{lemma:orthogonal}, we have $\nabla g(\bV)=\nabla g(\rho^{-1}\bV')=\nabla g(\bV')=\bm0$. The proof is then completed.
\end{IEEEproof}

We remark that problems \eqref{prob:MIMO:unconstrained} and \eqref{prob:MIMO} are not guaranteed to be equivalent since their global optimum may not be identical. Yet the two problems have the same set of stationary points excluding the zero solution. Because the beamforming problem is NP-hard \cite{luo2008dynamic}, it is adequate in most cases to reach a nonzero stationary point. As such, it suffices to consider the new problem \eqref{prob:MIMO:unconstrained} in the rest of the paper.

Next, we deal with problem \eqref{prob:MIMO:unconstrained} from an FP perspective. First, apply the Lagrangian dual transform \cite{shen2018fractional1} so as to move the matrix ratio $\bV_{k}^\hh\bH_{k}^\hh\widetilde\bF_{k}^{-1}
\bH_{k}\bV_{k}$ to the outside of log-det. As a result, problem \eqref{prob:MIMO:unconstrained} is converted to
\begin{subequations}
\label{prob:fr}
    \begin{align}
        \underset{\bV,\,\bGa}{\text{maximize}}\quad&\,f_r(\bV,\bGa)\\
\text{subject to}\quad
&\,\bGa_k\in\mathbb S_+,\; k=1,\ldots,K,
    \end{align}
\end{subequations}
where $\bGa=\{\bGa_k\}$ is a set of auxiliary variables, and the new objective function is given by
\begin{multline}
f_r(\bV,\bGa)=\sum_{k=1}^K\omega_{k}\Big[\log|\bI+\bGa_k|-\tr(\bGa_k)\\+\tr\big((\bI+\bGa_k)\bV_k^\hh\bH_{k}^\hh\bJ^{-1}_k\bH_k\bV_{k}\big)\Big] 
\end{multline}
with
\begin{align}
    \bJ_k &=\bH_{k}\bV_{k}\bV_{k}^\hh\bH_{k}^\hh + \widetilde\bF_k.
\end{align} 
When $\bV$ is held fixed, $\bGa$ in problem \eqref{prob:fr} can be optimally determined as
\begin{align}
\label{eq:update Ga}
\bGa_k^\star=\bV_k^{\hh}\bH_{k}^\hh\widetilde\bF_{k}^{-1}\bH_{k}\bV_{k}.
\end{align}
It remains to optimize $\bV$ when $\bGa$ is fixed. Note that problem \eqref{prob:fr} is a sum-of-matrix-ratios problem of $\bV$, for which the quadratic transform \cite{shen2018fractional1} can be readily applied. As a result, we further convert problem \eqref{prob:fr} to
\begin{subequations}
\label{prob:fq}
    \begin{align}
        \underset{\bV,\,{\bGa},\,\bY}{\text{maximize}}\quad&\,f_q(\bV,\bGa,\bY)\\
\text{subject to}\quad
&\,\bGa_k\in\mathbb S_+,\; k=1,\ldots,K\\
&\,\bY_k\in\mathbb C^{N\times d},\; k=1,\ldots,K,
    \end{align}
\end{subequations}
where $\bY=\{\bY_k\}$ is a set of auxiliary variables, and the new objective function is
\begin{align}
&f_q(\bV,\bGa,\bY)=\sum^K_{k=1}\omega_{k}\Big[\log|\bI+\bGa_k|-\tr(\bGa_k)\,+\notag
\\
&\;\tr\big(2\Re\{\bV_k^\hh\bH_k^\hh\bY_k(\bI+\bGa_k)\}-\bY_{k}^\hh\bJ_k\bY_{k}(\bI+\bGa_k)\big)\Big].
\end{align}
When $\bGa$ and $\bV$ are both held fixed, each $\bY_k$ can be optimally determined as
\begin{align}
\label{eq:update Y}
    \bY_{k}^\star = \bJ_k^{-1}\bH_{k}\bV_{k}.
\end{align}
We remark that both \eqref{eq:update Ga} and \eqref{eq:update Y} require inverting an $N\times N$ matrix. This shall not incur any computational difficulty since $N$ is a small number (e.g., $N=1$ or $2$ in a typical massive MIMO system \cite{rusek2012scaling}).

The key step is to optimize the precoding variable $\bV$ in \eqref{prob:fq} with $\bG$ and $\bY$ both held fixed. Note that $f_q(\bV,\bGa,\bY)$ is a concave quadratic function in $\bV$, so the solution of each $\bV_k$ can be obtained in closed-form by completing the square, i.e.,
\begin{align}
\label{eq:update V}
    \bV_{k}^\star=\bD^{-1}\bQ_k,
\end{align}
where
\begin{multline}
\label{eq:update D}
    \bD = \sum^K_{j=1}\omega_{j}\bigg[\bH_{j}^\hh\bY_{j}(\bI+\bGa_{j})\bY^\hh_{j}\bH_{j}\\+\frac{\sigma^2}{P}\tr(\bY_{j}^\hh\bY_{j}(\bI+\bGa_{j}))\bI\bigg],
\end{multline}
and 
\begin{align}
\label{eq:update Q}
    \bQ_k = \omega_k\bH_{k}^\hh\bY_{k}(\bI+\bGa_k),
\end{align}
As shown in \cite{shen2018fractional1}, updating $\bGa$, $\bY$, and $\bV$ iteratively guarantees convergence to a stationary point of \eqref{prob:fq}. We can subsequently recover a stationary point of the original problem \eqref{prob:fr} according to Theorem \ref{theorem:unconstrained}. Algorithm \ref{algorithm:WMMSE} summarizes the above steps; observe that it resembles the conventional WMMSE algorithm \cite{shi2011iteratively}.

Despite the closed-form steps in Algorithm \ref{algorithm:WMMSE}, 
the practical implementation of this algorithm can still be difficult because the optimal update of $\bV_k$ in \eqref{eq:update V} requires computing the $M\times M$ matrix inversion $\bD^{-1}$, where $M$ is a large number in our problem case. Thus, the real challenge is to optimize $\bV$ in \eqref{prob:fq} for fixed $(\bGa,\bY)$, which can be divided into a set of unconstrained quadratic program on a per $\bV_k$ basis:
\begin{align}
\underset{\bV_k}{\text{minimize}} &\quad \tr\Big(\frac12\bV_k^\hh\bD\bV_k-\Re\left\{\bV_k^\hh\bQ_k\right\}\Big),
\label{prob:QP}
\end{align}
for $k=1,2,\ldots,K$. The next section proposes a finite horizon optimization approach to the above quadratic program.

%To eliminate the $M\times M$ matrix inversion in \eqref{eq:update V}, a recent work \cite{zhang2023enhancing} suggests diagonalizing $\bD$ somehow, but it would slow down the convergence. A more recent work \cite{shen2024accelerating} can address this issue more or less by accelerating the convergence in the long run. In contrast, this work pursues a finite-horizon acceleration considering a limited number of iterations.

\begin{algorithm}[t]
  \caption{Unconstrained WMMSE Beamforming}
  \label{algorithm:WMMSE}
  \begin{algorithmic}[1]
      \STATE Initialize ${\bV}$ to any nonzero beamforming matrix.
      \REPEAT 
      \STATE Update each $\bm{\Gamma}_{k}$ by \eqref{eq:update Ga}.
      \STATE Update each $\bY_{k}$ by \eqref{eq:update Y}.
      \STATE Update each $\bV_k$ by \eqref{eq:update V}.
      \UNTIL{convergence} 
      \STATE  $\bV\leftarrow\rho\bV$, where 
    $\rho=(P/\sum^K_{k=1}\|\bV_k\|_F^2)^{1/2}$.
  \end{algorithmic}
\end{algorithm}

\section{Finite Horizon Optimization}

Denote by $g_{k}(\bV_{k})$ the objective function in \eqref{prob:QP}. To avoid the matrix inversion $\bD^{-1}$, we solve problem \eqref{prob:QP} by the gradient descent as
\begin{align}
\label{gradient}
&\bV_{k}^{t+1}=\bV_{k}^t-\eta_t \nabla g_{k}(\bV_{k}^t),
\end{align}
where $\bV^t_k$ is the beamforming variable decision in the $t$th iteration, $\eta_t\ge0$ is the step size, and the gradient can be computed as
\begin{equation}
 \nabla g_{k}(\bV_{k}^t) = \bD\bV_{k}^t - \bQ_k.
\end{equation}
In particular, to cut down the computation cost, we only allow the gradient update in \eqref{gradient} to run $T$ times, i.e., $t=0,1,\ldots,T-1$ in \eqref{gradient}. For the finite horizon optimization purpose, we seek the optimal step sizes $\eta_t$ to maximize the performance of the ultimate solution $\bV_{k}^T$ after $T$ iterations:
\begin{subequations}
\label{prob:finite gVk}
\begin{align}
\underset{\eta_0, \cdots, \eta_{T-1}}{\text{minimize}}& \quad g_{k}(\bV_{k}^T)\\
\text {subject to}& \quad \bV_{k}^{t+1}=\bV_{k}^t-\eta_t \nabla g_{k}(\bV_{k}^t).
\end{align} 
\end{subequations}
Denote by $\bV^\star_k$ the optimal solution to \eqref{prob:QP}. The objective function in \eqref{prob:finite gVk} can be rewritten as
\begin{equation}
   g_k(\bV_k) =\tr\Big( \frac{1}{2}(\bV_{k}-\bV_k^\star)^\hh\bD(\bV_{k}-\bV_k^\star)\Big)+C, 
\end{equation}
where $C$ is a constant term depending on $\bV_k^\star$. Moreover, let $\lambda_{M}$ be the largest eigenvalue of $\bD$ and $\lambda_{1}$ the smallest eigenvalue. Since $\bD$ is positive definite, we have $\lambda_{M}\ge\lambda_{1}>0$. It can be shown that
\begin{equation}
  \frac{\lambda_{1}}{2}\|\bV_k-\bV_k^*\|_F^2 \le g_k(\bV_k)-C \le \frac{\lambda_{M}}{2}\|\bV_k-\bV_k^*\|_F^2.
\end{equation}
Based on the above bounds, problem \eqref{prob:finite gVk} can be approximated as
\begin{subequations}
\label{prob:dist}
\begin{align}
\underset{\eta_0, \cdots, \eta_{T-1}}{\text{minimize}}&\quad \|\bV_k^T-\bV_k^*\|_F\\
\text {subject to}& \quad \bV_{k}^{t+1}=\bV_{k}^t-\eta_t \nabla g_{k}(\bV_{k}^t).
\end{align} 
\end{subequations}
Intuitively, we aim to minimize the (Euclidean) distance between the ultimate solution $\bV^T_k$ and the optimal solution $\bV^\star_k$ after $T$ iterations. 

Furthermore, we bound the distance $\|\bV_k^T-\bV_k^*\|_F$ from above as
\begin{align}
&\left\|\bV_{k}^T-\bV_k^\star\right\|_F \notag \\
& = \Bigg\|\left(
\bV_{k}^0-\bV_k^\star\right)\prod^{T-1}_{t=0}\left(\bI-\eta_{t} \bD\right)\Bigg\|_F\notag\\
 & \leq \Bigg\|\prod^{T-1}_{t=0}\left(\bI-\eta_{t} \bD\right)\Bigg\|_2\times\left\|\bV_{k}^0-\bV_k^*\right\|_F\notag \\
& \le \sup_{\lambda_{1}\le\lambda\le \lambda_{M}}\left|\prod^{T-1}_{t=0}\left(1-\eta_{t} \lambda\right)\right|\times\left\|\bV_{k}^0-\bV_k^*\right\|_F.
\label{dist upper bound}
\end{align}
Substituting this upper bound in problem \eqref{prob:dist}, we arrive at
\begin{align}
\label{prob:polydist}
\underset{\eta_0, \cdots, \eta_{T-1}}{\text{minimize}}&\quad \sup_{\lambda_{1}\le\lambda\le \lambda_{M}}\left|\prod^{T-1}_{t=0}\left(1-\eta_{t} \lambda\right)\right|.
\end{align} 
We shall decide the step sizes for the finite horizon optimization by solving the above problem.

\begin{proposition}
If we require all the step sizes to be identical, i.e., $\eta_0=\eta_1=\ldots=\eta_{T-1}=\eta$, then the optimal solution is
\begin{equation}
\label{constantstepsize}
    \eta = \frac{2}{\lambda_1+\lambda_M}.
\end{equation}
As a result, we obtain from \eqref{dist upper bound} that
\begin{equation}
\label{convergence:conventional GD}
\left\|\bV_{k}^T-\bV_k^\star\right\|_F\leq\left(1-\frac{2}{\kappa+1}\right)^T\left\|\bV_{k}^0-\bV_k^\star\right\|_F,
\end{equation}
where $\kappa=\lambda_M / \lambda_1$ is the condition number of matrix $\bD$.
\end{proposition}

Next, we consider \eqref{prob:polydist} with $(\eta_0, \cdots, \eta_{T-1})$ allowed to be different. It turns out that the problem is closely related to the degree-$T$ \emph{Chebyshev polynomial}:
\begin{align}
S(x) &= \cos(T\arccos x)\notag\\
&= 2^{T-1}(x-\xi_0)(x-\xi_1)\cdots(x-\xi_{T-1}),
\label{chebyshev}
\end{align}
where
\begin{equation}
\label{eq:xit}
    \xi_t = \cos \left(\left(t+\frac{1}{2}\right) \frac{\pi}{T}\right), \; t=0, 1, \ldots, T-1.
\end{equation}
For a constant $\gamma>1$, we normalize $S(x)$ at the position $\gamma$ as
\begin{equation}
    \widetilde S(x;\gamma) = \frac{S(x)}{S(\gamma)}.
\end{equation}
Clearly, we have $\widetilde S(\gamma;\gamma)=1$.

Now, an important result from \cite{markoff1916polynome,flanders1950numerical,young1953richardson} is that $\widetilde S(x;\gamma)$ has ``smaller'' fluctuations on the interval $[-1,1]$ than any other degree-$T$ polynomials that are normalized at $\gamma$, i.e.,
\begin{equation}
\label{chebyshov opt}
    \sup_{-1\le x \le 1}|\widetilde S(x;\gamma)| \le \sup_{-1\le x \le 1}|p(x)|,\quad\forall p(x)\in\mathcal P_\gamma,
\end{equation}
where
\begin{equation}
    \mathcal P_\gamma=\{p(x)\in\mathbb P_T: p(\gamma) =1\}.
\end{equation}
To utilize the above result, we need to rewrite the objective function of \eqref{prob:polydist} in the form of the Chebyshev polynomial. Denote by $S_0(\lambda)$ the objective function in \eqref{prob:polydist} as a polynomial of $\lambda$ given $(\eta_0, \cdots, \eta_{T-1})$:
\begin{equation}
    S_0(\lambda) = \prod^{T-1}_{t=0}\left(1-\eta_{t} \lambda\right).
\end{equation}
Letting
\begin{equation}
\label{x}
    x = \frac{\lambda_M+\lambda_1}{\lambda_M-\lambda_1} - \frac{2}{\lambda_M-\lambda_1} \lambda
\end{equation}
so that $-1\le x\le 1$ when $\lambda_1\le \lambda \le \lambda_M$, we rewrite $S_0(\lambda)$ as a polynomial of $x$:
\begin{equation}
    S_1(x) = \prod^{T-1}_{t=0}\left(1-\eta_{t}\left(\frac{\lambda_M+\lambda_1}{2}-\frac{\lambda_M-\lambda_1}{2}x\right)\right).
\end{equation}
Because of the one-to-one correspondence between $x$ and $\lambda$, we have the problem equivalence:
\begin{equation*}
    \underset{\eta_0, \cdots, \eta_{T-1}}{\text{minimize}} \sup_{\lambda_{1}\le\lambda\le \lambda_{M}} \big|S_0(\lambda)\big| \;\Longleftrightarrow\;
        \underset{\eta_0, \cdots, \eta_{T-1}}{\text{minimize}} \sup_{-1\le x\le 1} \big|S_1(x)\big|.
\end{equation*}
Observing that every possible $S_1(x)$ equals 1 when $x=\frac{\lambda_M+\lambda_1}{\lambda_M-\lambda_1}$, we further have
\begin{equation*}
    \underset{\eta_0, \cdots, \eta_{T-1}}{\text{minimize}} \sup_{-1\le x\le 1} \big|S_1(x)\big| \;\Longleftrightarrow\;
        \underset{p(x)\in\mathcal P_\gamma}{\text{minimize}} \sup_{-1\le x\le 1} \big|p(x)\big|,
\end{equation*}
where
\begin{equation}
    \gamma = \frac{\lambda_M+\lambda_1}{\lambda_M-\lambda_1}.
\end{equation}
According to \eqref{chebyshov opt}, the optimal solution of $(\eta_0,\ldots,\eta_{T-1})$ is the one that makes
\begin{equation}
    S_1(x) = \widetilde S\bigg(x;\frac{\lambda_M+\lambda_1}{\lambda_M-\lambda_1}\bigg).
\end{equation}
In other words, we should choose $(\eta_0,\ldots,\eta_{T-1})$ to let $S_1(x)S(\frac{\lambda_M+\lambda_1}{\lambda_M-\lambda_1})$ be the degree-$T$ Chebyshev polynomial. This optimal solution is stated in the following proposition.
\setcounter{equation}{52}
\begin{figure*}[b]
\hrule
\begin{align}
\label{multicell:fq}
f^{\mathrm{MC}}_q(\bV,\bGa,\bY)=\sum_{\ell=1}^L\sum^K_{k=1}\omega_{\ell k}\Big[\log|\bI+\bGa_{\ell k}|-\tr(\bGa_{\ell k})
+\tr\big(2\Re\{\bV_{\ell k}^\hh\bH_{\ell k,\ell}^\hh\bY_{\ell k}(\bI+\bGa_{\ell k})\}
-\bY_{\ell k}^\hh\bJ_{\ell k}\bY_{\ell k}(\bI+\bGa_{\ell k})\big)\Big]
\end{align}
\end{figure*}

\begin{proposition}
\label{prop:opt_eta}
An optimal solution to problem \eqref{prob:polydist} is
\setcounter{equation}{43}
\begin{equation}
\label{eq:stepsizesingle}
\eta_t^\star=\bigg(\frac{\lambda_M+\lambda_1}{2}+\frac{\lambda_M-\lambda_1}{2}\xi_t\bigg)^{-1},
\end{equation}
for $t=0,1,\ldots,T-1$. This optimal solution leads to
\begin{equation}
\label{convergence:finite}
\left\|\bV_{k}^T-\bV_k^\star\right\|_F\leq2\left(1-\frac{2}{\sqrt{\kappa}+1}\right)^T\left\|\bV_{k}^0-\bV_k^\star\right\|_F,
\end{equation}
where $\kappa=\lambda_M / \lambda_1$ is the condition number of matrix $\bD$.
\end{proposition}

\begin{remark}
    The convergence bound in \eqref{convergence:conventional GD} continues to hold when $T$ is replaced by any other positive integer $t=1,2,\ldots$, whereas the convergence bound in \eqref{convergence:finite} holds only for this particular $T$.
\end{remark}

\begin{remark}
    Problem \eqref{prob:polydist} has multiple optimal solutions. Actually, since the variables $\eta_0,\eta_1,\ldots,\eta_{T-1}$ are symmetric in the objective function of \eqref{prob:polydist}, the reordering of $(\eta^\star_0,\eta^\star_1,\ldots,\eta^\star_{T-1})$ in \eqref{eq:stepsizesingle} still gives an optimal solution.
\end{remark}

%\begin{remark}
%The order of the stepsizes $\{\eta_0,\eta_1,\ldots,\eta_{T-1}\}$ can be arbitrarily permuted without altering the value of the objective function at the end of the algorithm.
%\end{remark}

%\begin{remark}
%The complexity of the aforementioned stepsize rules only holds at $T$-th step but not at any other iterations, so it does not have asymptotic guarantees like constant stepsize.
%\end{remark}
    
\begin{algorithm}[t]
  \caption{Finite Horizon Beamforming for a Single Cell}
  \label{algorithm:Finite FP single}
  \begin{algorithmic}[1]
      \STATE Initialize ${\bV}$ to any nonzero beamforming matrix.
      \REPEAT 
      \STATE Update each $\bm{\Gamma}_{k}$ by \eqref{eq:update Ga}.
      \STATE Update each $\bY_{k}$ by \eqref{eq:update Y}.
      \STATE Update each $\bV_k$ by \eqref{gradient} for $T$ iterations, with the step sizes determined as in \eqref{eq:stepsizesingle}.
      \UNTIL{convergence} 
      \STATE  $\bV\leftarrow\rho\bV$, where 
    $\rho=(P/\sum^K_{k=1}\|\bV_k\|_F^2)^{1/2}$.
  \end{algorithmic}
\end{algorithm}

\section{Multi-Cell Case}

Thus far our attention is limited to the single-cell downlink network with only one BS. This section aims at a multi-cell extension. Assume that there are $L$ BSs, each equipped with $M$ transmit antennas. The rest setting follows that in the single-cell case, e.g., each cell has $K$ user terminals, and each user terminal has $N$ receive antennas. We use $\ell,i=1,\ldots,L$ to index the cells and the corresponding BSs, and still use $k,j=1,\ldots,K$ to index the users within each cell. Denote by $\bH_{\ell k,i}\in\mathbb C^{N\times M}$ the channel from BS $i$ to the $k$th user in cell $\ell$, and $\bV_{\ell k}\in\mathbb C^{M \times d}$ the transmit precoder of BS $\ell$ for its $k$th associated user. As a result, the achievable rate of the $k$th user in cell $\ell$, written $R_{\ell k}$, is given by
\begin{equation}
\label{multirate}
R_{\ell k} = \log|\bI+\bV_{\ell k}^\hh\bH_{\ell k,\ell}^\hh\bF_{\ell k}^{-1}
\bH_{\ell k,\ell}\bV_{\ell k}|,
\end{equation}
where
\begin{equation}
   \label{Flk}
   \bF_{\ell k} = \sigma^2\bI+\sum_{(i,j)\ne (\ell,k)}\bH_{\ell k,i}\bV_{ij}\bV_{ij}^\hh\bH_{\ell k,i}^\hh.
\end{equation}

The weighted sum-of-rates maximization problem across multiple cells is:
\begin{subequations}
\label{prob:multMIMO}
\begin{align}
\underset{\bV}{\text{maximize}} &\quad \sum^L_{\ell=1}\sum^K_{k=1}\omega_{\ell k}R_{\ell k}
\label{prob:multMIMO:obj}\\
  \text{subject to} & \quad \sum^K_{k=1}\|\bV_{\ell k}\|_F^2\leq P,\; \ell=1,\ldots,L.
  \label{prob:multMIMO:constraint}
\end{align}
\end{subequations}
Unlike the single-cell case, the power constraint in the above problem need not be tight at the optimum point. We propose rewriting \eqref{prob:multMIMO} as
\begin{subequations}
\label{prob:multMIMO:new}
\begin{align}
\underset{\bV}{\text{maximize}} &\quad \sum^L_{\ell=1}\sum^K_{k=1}\omega_{\ell k}R_{\ell k}
\label{prob:multMIMO:new:obj}\\
  \text{subject to} & \quad \sum^K_{k=1}\|\bV_{\ell k}\|_F^2= P_\ell,\; \ell=1,\ldots,L,
  \label{prob:multMIMO:new:constraint}
\end{align}
\end{subequations}
with the power variables $P_\ell\le P$. In practice, we may optimize $P_\ell$ separately via the existing power control method for fixed $\bV$.

Repeating the procedure from \eqref{prob:MIMO:tight} to \eqref{prob:MIMO:unconstrained} in Section \ref{sec:QP}, we recast problem \eqref{prob:multMIMO:new} into an unconstrained problem:
\begin{align}
\underset{\bV}{\text{maximize}} &\quad \sum_{\ell,k}\omega_{\ell k} \log\big|\bI+\bV_{\ell k}^\hh\bH_{\ell k,\ell}^\hh\widetilde{\bF}_{\ell k}^{-1}
\bH_{\ell k,\ell}\bV_{\ell k}\big|
\label{prob:MIMO:obj:newmult}
\end{align}
where
\begin{equation}
    \label{multFlk}
   \widetilde\bF_{\ell k} = \Bigg(\frac{\sigma^2}{P_\ell}\sum^K_{k=1}\|\bV_{\ell k}\|^2_F\Bigg)\bI+\sum_{(i,j)\ne (\ell,k)}\bH_{\ell k,i}\bV_{ij}\bV_{ij}^\hh\bH_{\ell k,i}^\hh.
\end{equation}
Again, by the matrix FP technique, we obtain the further reformulation as an unconstrained quadratic program:
\begin{subequations}
\label{prob:multfq}
    \begin{align}
        \underset{\bV,\,{\bGa},\,\bY}{\text{maximize}}\quad&\,f^{\mathrm{MC}}_q(\bV,\bGa,\bY)\\
\text{subject to}\quad
&\,\bGa_{\ell k}\in\mathbb S_+,\; \forall(\ell,k)\\
&\,\bY_{\ell k} \in\mathbb C^{N\times d},\;\forall(\ell,k),
    \end{align}
\end{subequations}
where the new objective function is shown in 
\eqref{multicell:fq} along with 
\setcounter{equation}{53}
\begin{align}
    \bJ_{\ell k} &=\bH_{\ell k,\ell}\bV_{\ell k}\bV_{\ell k}^\hh\bH_{\ell k,\ell}^\hh + \widetilde\bF_{\ell k}.
\end{align}
Similar to the single-cell case, the two auxiliary variables are optimally determined as
\begin{align}
\bGa_{\ell k}^\star&=\bV_{\ell k}^{\hh}\bH_{\ell k,\ell}^\hh\widetilde\bF_{\ell k}^{-1}\bH_{\ell k,\ell}\bV_{\ell k},\label{eq:multupdate Ga}\\
\bY_{\ell k}^\star &= \bJ_{\ell k}^{-1}\bH_{\ell k,\ell}\bV_{\ell k}.
\label{eq:multupdate Y}
\end{align}
\begin{algorithm}[t]
  \caption{Finite Horizon Beamforming for Multiple Cells}
  \label{algorithm:Finite FP multi}
  \begin{algorithmic}[1]
      \STATE Initialize ${\bV}$ to any nonzero beamforming matrix.
      \REPEAT 
      \STATE Update each $\bm{\Gamma}_{\ell k}$ by \eqref{eq:multupdate Ga}.
      \STATE Update each $\bY_{\ell k}$ by \eqref{eq:multupdate Y}.
      \STATE Update each $\bV_k$ by \eqref{multicell:gradient} for $T$ iterations, with the step sizes determined as in \eqref{eq:stepsizemultiple}.
      \UNTIL{convergence} 
      \STATE  $\bV_{\ell k}\leftarrow\rho_\ell\bV_{\ell k}$, where 
    $\rho_\ell=(P_\ell/\sum^K_{k=1}\|\bV_ {\ell k}\|_F^2)^{1/2}$.
  \end{algorithmic}
\end{algorithm}

For fixed auxiliary variables, we update the beamforming variable by gradient descent up to $T$ iterations:
\begin{align}
\label{multicell:gradient}
&\bV_{\ell k}^{t+1}=\bV_{\ell k}^t-\eta_{\ell,t} \nabla g_{\ell k}(\bV_{k}^t),
\end{align}
%where 
%\begin{align}
%    g_{\ell k}(\bV_{\ell k}) = \tr\Bigg(\bV_{\ell k}^\hh\bD_{\ell }\bV_{\ell k}-2\Re\{\omega_{\ell k}\bV_{\ell k}^\hh\bH_{\ell k,\ell}^\hh\bY_{\ell k}(\bI+\bGa_{\ell k})\}\Bigg)
%\end{align}
where
\begin{align}
    \nabla g_{\ell k}(\bV_{\ell k}) = \bD_\ell \bV_{\ell k} -\omega_{\ell k}\bH_{\ell k,\ell}^\hh\bY_{\ell k}(\bI+\bGa_{\ell k})
\end{align}
Now, the step sizes $(\eta_{\ell,0},\eta_{\ell,1},\ldots,\eta_{\ell,T-1})$ are determined for each cell $\ell$ individually as in Proposition \ref{prop:opt_eta}:
\begin{equation}
\label{eq:stepsizemultiple}
\eta_{\ell,t}=\bigg(\frac{\lambda_{\ell M}+\lambda_{\ell 1}}{2}+\frac{\lambda_{\ell M}-\lambda_{\ell 1}}{2}\xi_t\bigg)^{-1},
\end{equation}
only that the largest eigenvalue $\lambda_{\ell M}$ and the smallest eigenvalue $\lambda_{\ell 1}$ are now associated with the cell-specific matrix
\begin{multline}
   \bD_{\ell} = \sum^L_{i=1}\sum^K_{j=1}\omega_{ij}\bH_{ij,\ell}^\hh\bY_{ij}(\bI+\bGa_{ij})\bY^\hh_{ij}\bH_{ij,\ell}\\
    +\sum^K_{j=1}\frac{\omega_{\ell j} \sigma^2}{P_\ell}\tr\big(\bY_{\ell j}^\hh\bY_{\ell j}(\bI+\bGa_{\ell j})\big)\bI.
    \label{eq:multupdate D}
\end{multline}
Algorithm \ref{algorithm:Finite FP multi} summarizes the above steps for the extended finite horizon beamforming for multiple cells.

\section{Simulation Results} 
We first validate the performance of the proposed algorithms numerically in a single-cell large-scale MIMO network. Within the cell, the BS is located at the center, and $6$ downlink users are randomly distributed. Let the BS have $2048$ transmit antennas, and let each downlink user have $8$ receive antennas, $8$ independent data streams being intended for each downlink receiver. The maximum transmit power $P$ equals $20$ dBm. The background noise level $\sigma^2$ equals $-80$ dBm. The rate weights $\omega_k$ are all set to $1$. The downlink distance-dependent path-loss is modeled as $15.3 + 37.6 \log_{10}(l) + \Delta $ (in dB), where $l$ represents the distance in meters, and $\Delta$ is a zero-mean Gaussian random variable with a standard variance of $8$ dB---which models the shadowing effect. The parameter $T$ is fixed at $5$ for Algorithm \ref{algorithm:Finite FP single} by default. Aside from Algorithm \ref{algorithm:WMMSE} and Algorithm \ref{algorithm:Finite FP single}, we consider the conventional gradient descent with equal step size as in \eqref{constantstepsize}; we let the conventional gradient descent run $T$ iterations as well for fairness.
%H. Lu and Y. Zeng, “Communicating with Extremely LargeScale Array/ Surface: Unified Modeling and Performance Analysis,” IEEE Trans. Wireless Commun., vol. 21, no. 6, June 2022, pp. 4039–53.

\begin{figure}[t]
    \centering
    \includegraphics[width=1\linewidth]{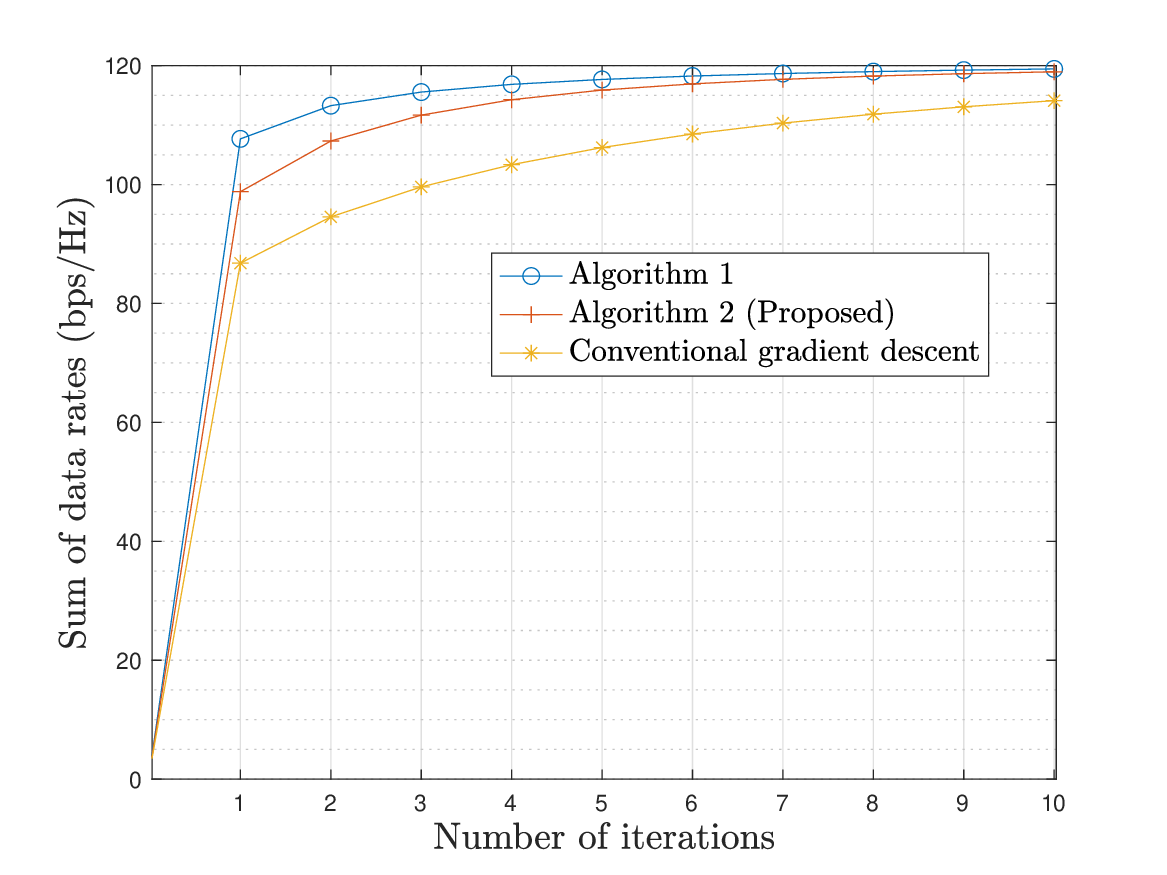}
    \caption{Sum rates vs. number of iterations in the single-cell case.}
    \label{fig1}
\end{figure}

\begin{figure}[t]
    \centering
    \includegraphics[width=1\linewidth]{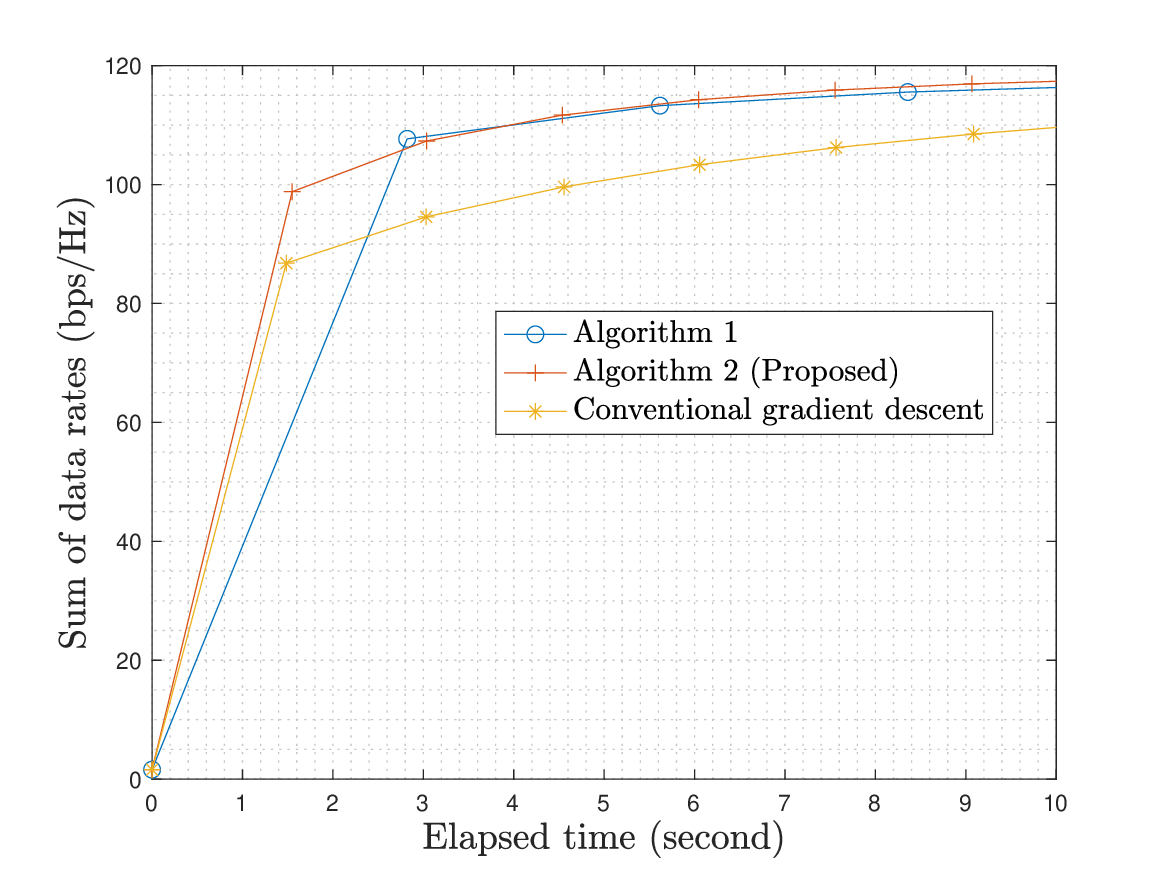}
    \caption{Sum rates vs. CPU time in the single-cell case.}
    \label{fig2}
\end{figure}

Fig.~\ref{fig1} and Fig.~\ref{fig2} compare the convergences of the different beamforming algorithms in iterations and in CPU time, respectively. Observe from Fig.~\ref{fig1} that Algorithm \ref{algorithm:WMMSE}, i.e., the modified WMMSE algorithm without need to tune the Lagrangian multiplier for the power constraint, converges faster than the proposed Algorithm \ref{algorithm:Finite FP single} and conventional gradient descent algorithm in iterations. This is because Algorithm \ref{algorithm:WMMSE} finds the exact global optimum of the quadratic program in \eqref{prob:QP} every after the updates of the auxiliary variables. From an MM perspective, it gives a tighter approximation of the original objective function \eqref{prob:MIMO}. However, this is at a high cost of computing a $2048\times2048$ matrix inversion for each iteration. As a result, it can be seen from Fig.~\ref{fig2} that the proposed Algorithm \ref{algorithm:Finite FP single} converges faster than Algorithm \ref{algorithm:WMMSE} in CPU time, since the per-iteration complexity of the former is much lower than that of the latter. For instance, Algorithm \ref{algorithm:WMMSE} requires about $2.6$ seconds to reach $100$ bps/Hz, while the proposed Algorithm \ref{algorithm:Finite FP single} merely requires $1.6$ seconds to attain the same performance.

\begin{figure}[t]
    \centering
    \includegraphics[width=1\linewidth]{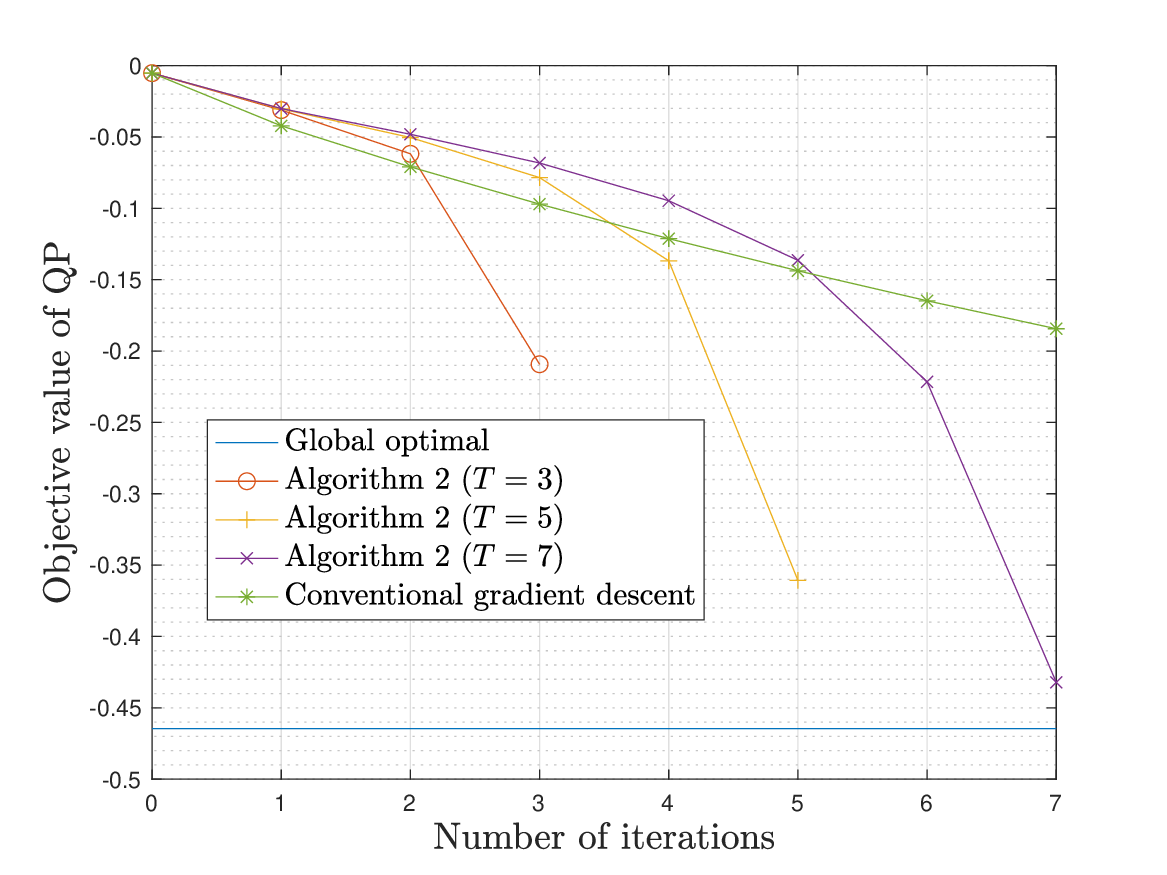}
    \caption{Objective value in \eqref{prob:QP} vs. number of iterations in the single-cell case.}
    \label{fig3}
\end{figure}

Furthermore, Fig.~\ref{fig3} focuses on how the different algorithms tackle the quadratic program in \eqref{prob:QP} iteratively when the auxiliary variables $\bGa_k$ and $\bY_k$ are both held fixed. Note that Algorithm \ref{algorithm:WMMSE} is not included in the figure because it solves the quadratic program \eqref{prob:QP} directly. We try out different values of $T$ for Algorithm \ref{algorithm:Finite FP single}, i.e., $T\in\{3,5,7\}$. According to Fig.~\ref{fig3}, Algorithm \ref{algorithm:Finite FP single} converges more quickly when $T$ is smaller, but its ultimate performance after $T$ iterations becomes worse. Observe also that Algorithm \ref{algorithm:Finite FP single} need not be always faster than the conventional gradient descent, but when it terminates at the $T$th iteration, it must outperform the conventional gradient descent. Further, we remark that Algorithm \ref{algorithm:Finite FP single} is already quite close to the global optimum when $T=7$, so the quadratic program in \eqref{prob:QP} can be well addressed without computing the large matrix inversion. 

%when applying different algorithms. As illustrated in Fig.~\ref{fig3}, the conventional gradient descent algorithm maintains a nearly consistent convergence slope over 7 iterations. This is due to its constant step size mechanism. For the proposed Algorithm \ref{algorithm:Finite FP single}, we systematically consider cases with $T = \{3, 5, 7\}$. The convergence results show the behavior: larger $T$ induces slower initial descent rates but achieves accelerated terminal-phase convergence. This characteristic is fundamentally governed by the spectral root distribution of Chebyshev polynomials, where larger $T$ amplifies the terminal iteration's effective step size through polynomial root displacement. Moreover, our theoretical analysis demonstrates that $T > 10$ may lead to numerical instabilities that paradoxically increase the objective function value. 

\begin{figure}[t]
    \centering
    \includegraphics[width=1\linewidth]{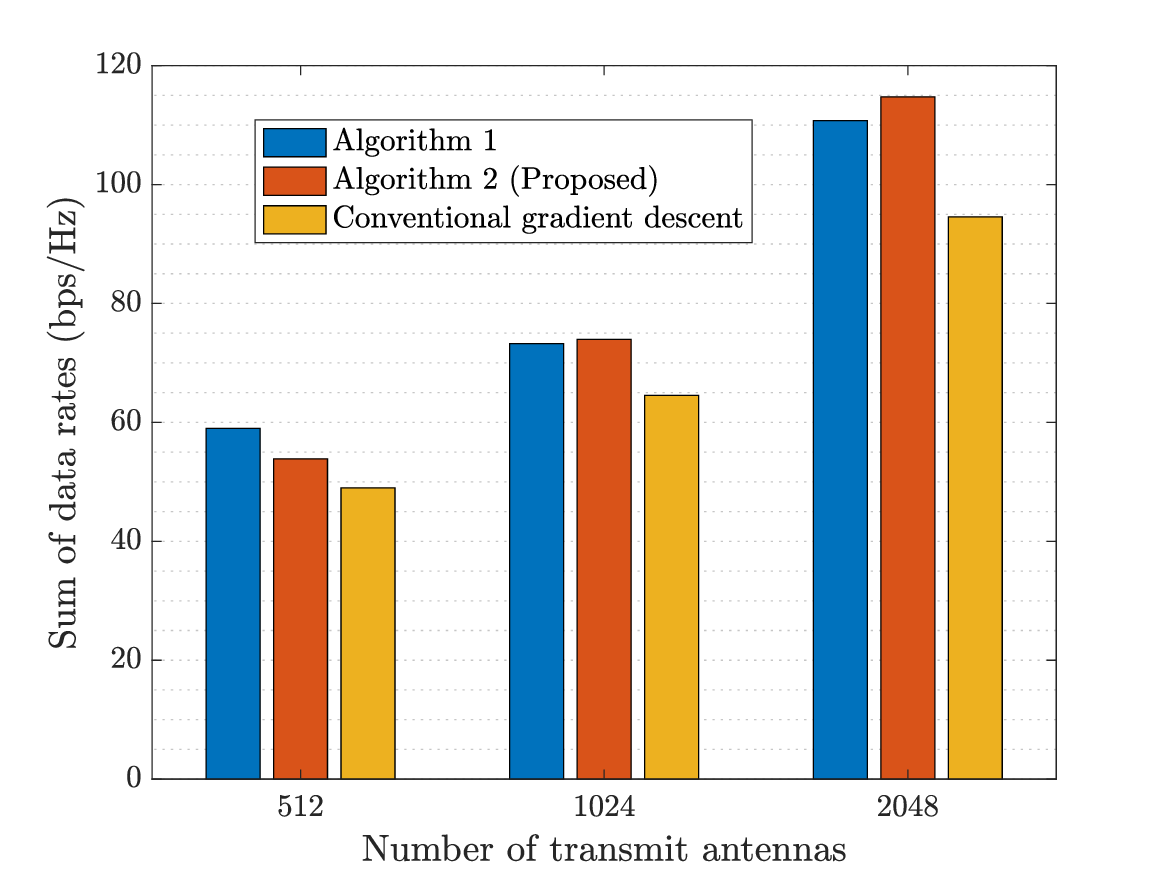}
    \caption{Sum rates vs. number of transmit antennas in the single-cell case.}
    \label{fig4}
\end{figure}

Next, Fig.~\ref{fig4} shows how the different algorithms behave when the number of transmit antennas $M$ is changed. For fair comparison, the same amount of CPU time is given to all the algorithms. Clearly, Algorithm \ref{algorithm:WMMSE}, which is the modified WMMSE algorithm, is sensitive to the increase of the number of antennas. While Algorithm \ref{algorithm:WMMSE} outperforms Algorithm \ref{algorithm:Finite FP single} when $M=512$, it is surpassed when $T$ increases to $1024$. The benefit of eliminating the large matrix inversion now starts to pay off. The advantage of Algorithm \ref{algorithm:Finite FP single} over Algorithm \ref{algorithm:WMMSE} becomes even larger when $T$ further increases to $2048$.

%compares the performance of three algorithms under fixed computational time constraints with different numbers of transmit antennas $M$. The proposed Algorithm \ref{algorithm:Finite FP single} exhibits $M$-scalable superiority: its performance transitions from inferiority relative to Algorithm 1 at $M = 512 $, achieves parity at $M = 1024$, and ultimately demonstrates dominance at $M = 2048$. This phenomenon arises from fundamentally distinct computational complexity regimes: Algorithm \ref{algorithm:WMMSE} suffers $\mathcal{O}(M^3)$ complexity dominated by matrix inversion in variable $\bV$'s updates, whereas the proposed Algorithm \ref{algorithm:Finite FP single} reduces this to $\mathcal{O}(M^2)$ through eigenvalue extraction of matrix $\bD$ via power iteration. The results establish the proposed Algorithm \ref{algorithm:Finite FP single} as a complexity-scalable solution for large-scale antenna systems.

We now consider the multi-cell case. The different algorithms are tested in a 3-cell large-scale MIMO network, with the BS-to-BS distance set to $800$ meters. The rest settings (e.g., the number of transmit antennas of each BS) follows the previous single-cell case. We do not consider power control in this case, simply letting each $P_\ell=P$. 

\begin{figure}[t]
    \centering
    \includegraphics[width=1\linewidth]{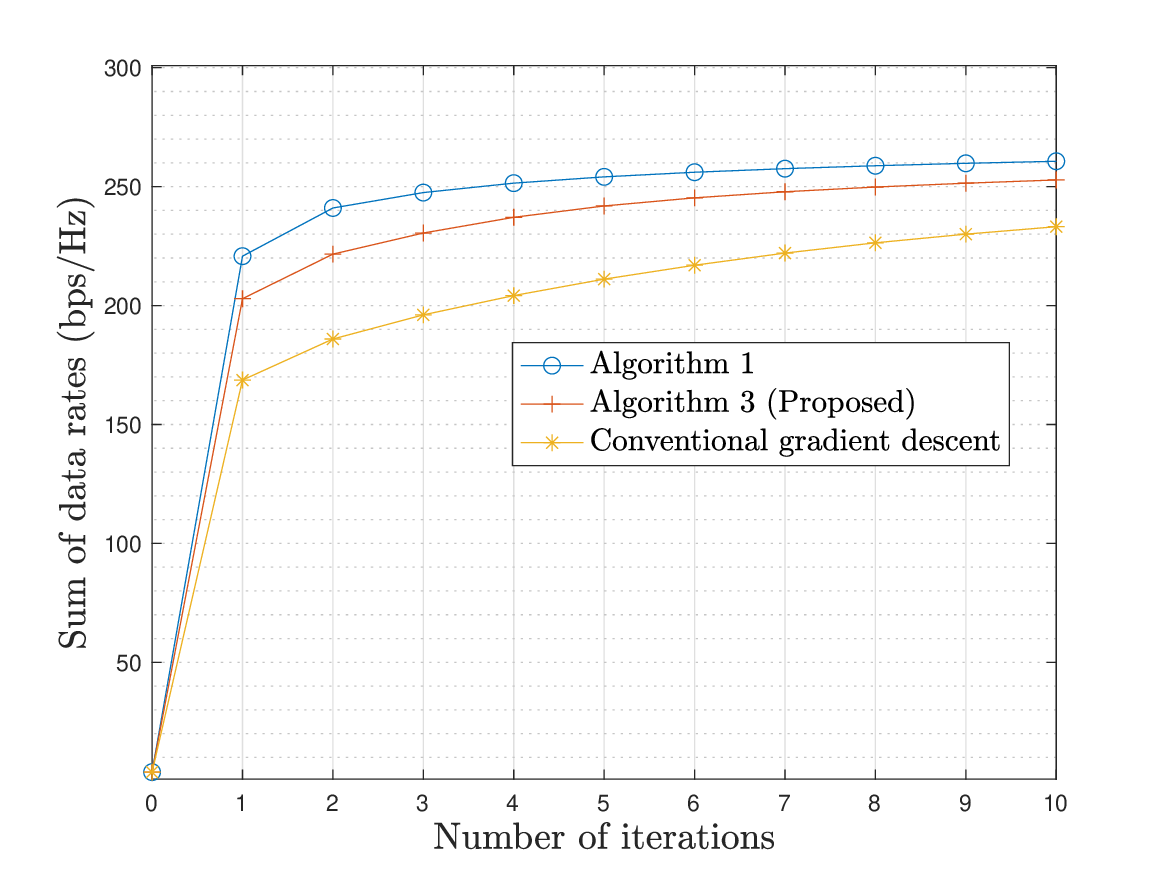}
    \caption{Sum rates vs. number of iterations in the multi-cell case.}
    \label{fig5}
\end{figure}

\begin{figure}[t]
    \centering
    \includegraphics[width=1\linewidth]{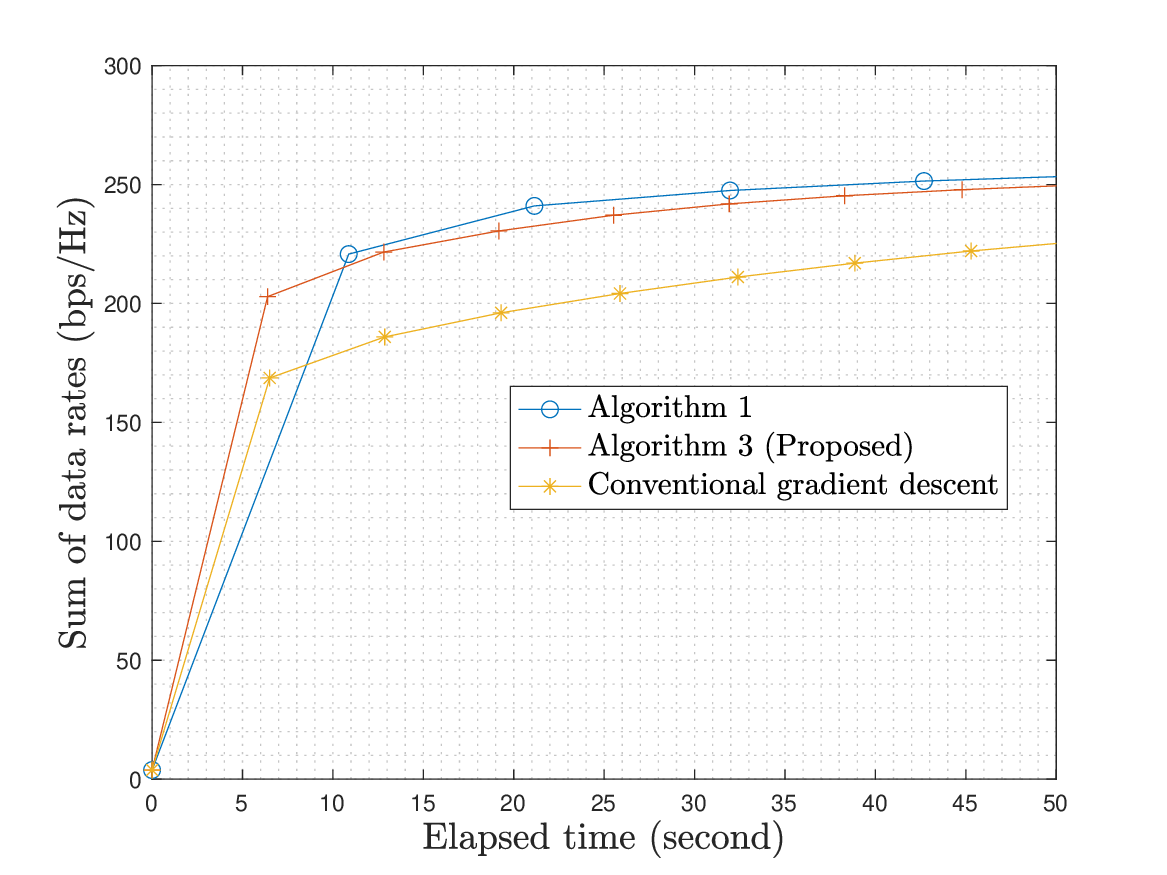}
    \caption{Sum rates vs. CPU time in the multi-cell case.}
    \label{fig6}
\end{figure}

Fig.~\ref{fig5} and Fig.~\ref{fig6} repeat the convergence tests for the multi-cell network. The results do not differ much from the previous single-cell case. Again, the modified WMMSE of Algorithm \ref{algorithm:WMMSE} requires the fewest iterations to converge according to Fig.~\ref{fig5}, and yet is much slower than the proposed Algorithm \ref{algorithm:Finite FP single} according to Fig.~\ref{fig6} when it comes to CPU time, because of its high computation cost in large matrix inversion. Actually, Algorithm \ref{algorithm:WMMSE} is even slower than the conventional gradient descent as shown in Fig.~\ref{fig6}. For instance, Algorithm \ref{algorithm:WMMSE} requires almost twice as much time as the other two algorithms to finish the first iteration.

%illustrate the convergence behavior of the different algorithm in multi-cell cases. As demonstrated, the multi-cell scenario is analogous to the single-cell case. Fig.~\ref{fig5} shows that Algorithm \ref{algorithm:WMMSE} still outperforms \ref{algorithm:Finite FP single} in terms of convergence in number of iterations. Regarding runtime performance, Algorithm \ref{algorithm:Finite FP single} achieves significantly better results in its first iteration compared to Algorithm \ref{algorithm:WMMSE}, reaching a rate of 200 within 6 seconds, whereas Algorithm \ref{algorithm:WMMSE} requires 10 seconds. However, in subsequent iterations, Algorithm \ref{algorithm:Finite FP single} does not exhibit the same advantages as in the single-cell scenario. This may be attributed to the fact that our new problem formulation cannot be strictly equivalent to the original problem in the multi-cell context.

\section{Conclusion}

This paper proposes a novel finite-horizon approach to the beamforming problem for the large-scale MIMO. We first convert the constrained log-det problem to an unconstrained quadratic program by the matrix FP. Optimizing the beamforming variables in the new problem iteratively by the gradient descent can avoid the costly matrix inversion. Our key contribution is to incorporate finite horizon optimization into the gradient descent to enhance efficiency. This is the very first application case of finite horizon optimization in the area of signal processing and digital communications.

\bibliographystyle{IEEEtran}
\bibliography{IEEEabrv,refs}
\end{document}